\documentclass[floatfix,aps,prd,twocolumn,letterpaper,lengthcheck,superscriptaddress,showpacs,amssymb,amsmath,amsfonts,nofootinbib,altaffilletter,nopreprintnumbers,showpacs,longbibliography]{revtex4-1}

\usepackage[utf8]{inputenc}
\usepackage{xspace}
\usepackage{acronym}
\usepackage{booktabs}
\usepackage{amsmath,amssymb,graphicx}
\usepackage{color}
\usepackage[colorlinks, linkcolor=blue, pdfborder={0 0 0}, breaklinks=true]{hyperref}
\usepackage[table]{xcolor}
\definecolor{lightgray}{gray}{0.9} 
\usepackage{tabularx}
\usepackage{multirow}
\usepackage{bigstrut}
\usepackage{braket}
\usepackage{makecell}
\usepackage{mathtools}
\usepackage{lineno}
\usepackage{aas_macros}

\DeclareUnicodeCharacter{2009}{\,}

\DeclareMathOperator{\Tr}{Tr}
\newcommand{\norm}[1]{\left\lVert#1\right\rVert}

\begin{document}
	
\title{Signal Space in the Triangular Network of Einstein Telescope}

\author{Isaac C. F.~Wong}
\email{cfwong@link.cuhk.edu.hk}
\affiliation{Department of Physics, The Chinese University of Hong Kong, Shatin, N.T., Hong Kong}
\author{Tjonnie G. F.~Li}
\affiliation{Department of Physics, The Chinese University of Hong Kong, Shatin, N.T., Hong Kong}
\affiliation{Institute for Theoretical Physics, KU Leuven, Celestijnenlaan 200D, B-3001 Leuven, Belgium}
\affiliation{Department of Electrical Engineering (ESAT), KU Leuven, Kasteelpark Arenberg 10, B-3001 Leuven, Belgium}

\begin{abstract}
	\noindent The proposed third-generation gravitational-wave detectors Einstein Telescope will have a triangular design that consists of three colocated interferometers. Summing the strain outputs from the three interferometers will cancel any gravitational-wave signal and the resultant signal-free stream is known as null stream. The null stream is in a fixed subspace of the observation space of Einstein telescope where no gravitational-wave signal can exist. In this paper, we establish the decomposition of the observation space of Einstein Telescope into the signal space that contains all possible gravitational-wave signals and the null space that contains the null stream. We show that the results of Bayesian parameter estimation and model selection using the strain data in the signal space are identical to that using the full set of strain data. This implies that one could use a fraction of the strain data to extract all information of the source which reduces the memory usage and speeds up the likelihood evaluation. We also discuss the existence of a fixed null space in Einstein Telescope allows the unbiased estimation of the noise properties in the signal-free subspace.
\end{abstract}

\maketitle

\section{Introduction}

\noindent The second-generation (2G) ground-based gravitational-wave (GW) detectors, Advanced LIGO \cite{2015} and Advanced Virgo \cite{Acernese_2014}, have detected dozens of GW events \cite{PhysRevX.9.031040,Abbott_2021} since the first detection of the GW event from a binary black hole coalescence in 2015 \cite{PhysRevLett.116.061102}. The observation of GWs opens a new window to study the Universe. The proposed third-generation (3G) ground-based GW detectors Einstein Telescope (ET) \cite{Punturo_2010,Hild_2011} will further help to address numerous key issues in astrophysics, fundamental physics, and cosmology thanks to its improved sensitivity by an order of magnitude \cite{Maggiore_2020}.

In contrast to the L-shaped interferometers with orthogonal arms in Advanced LIGO and Advanced Virgo, ET is proposed with a triangular design that consists of three colocated interferometers with the arms forming a $60^{\circ}$ angle \cite{Punturo_2010}. The ET configuration allows the construction of a null stream independent of the sky location of the source. The null stream in ET is suggested to be a powerful tool to estimate the power spectral density (PSD) of detector noise and vetoing non-Gaussian noise transients \cite{Regimbau_2012}.

In the literature \cite{Regimbau_2012,Maggiore_2020}, null stream in ET is simply regarded as the sum of the strain outputs from the three interferometers which cancels any GW signal, and it is not well recognized that null stream is in fact in a fixed subspace of the observation space where no GW can exist. The whole observation space can be decomposed into two orthogonal subspaces which we term as the signal space and the null space respectively. While the existence of null stream in ET is well discussed in the literature, there is no discussion of the complement subspace i.e.\ the signal space. The complement signal space contains all possible GW signals and this implies that we can extract all GW information from a fraction of the strain data. In this paper, we will establish the decomposition of the observation space into the signal space and the null space.

The paper is structured as follows. In Sec.~\ref{sec:det_response}, we discuss the existence of a null stream independent of the source location in ET. In Sec.~\ref{sec:signal_space}, we present the decomposition of the observation space of ET into the signal space and the null space. In Sec.~\ref{sec:pe}, we discuss performing parameter estimation with the signal-space strain data and introduce the corresponding likelihood functions in the time domain and frequency domain respectively. The paper is summarized in Sec.~\ref{sec:conclusion}.

\section{Detector response}
\label{sec:det_response}

\noindent With the long wavelength approximation, the detector response of a GW can be written as \cite{anderson_2011}
\begin{equation}
	\label{eq:det_response}
	\begin{split}
		s(t) &= \sum_{m}\sum_{i,j}D_{ij}e_{m}^{ij}h_{m}(t) \\
		&= \sum_{m}\Tr(\boldsymbol{D}\boldsymbol{e}_{m})h_{m}(t)
	\end{split}
\end{equation}
where $D_{ij}$ is the detector tensor that is determined by the geometry of the detector, $e_{m}^{ij}$ is the symmetric polarization tensor of mode $m$, and $h_{m}(t)$ is the polarization mode $m$. The detector tensor of a detector is given by
\begin{equation}
	\boldsymbol{D} = \frac{1}{2} \left( \boldsymbol{u}\otimes\boldsymbol{u} - \boldsymbol{v}\otimes\boldsymbol{v} \right)
\end{equation}
or
\begin{equation}
	D_{ij} = \frac{1}{2} \left( u_{i}u_{j} - v_{i}v_{j} \right)
\end{equation}
where $\boldsymbol{u}$ and $\boldsymbol{v}$ are the unit vectors along the two arms of the interferometer, and $\otimes$ is the outer product. The triangular configuration of ET would imply the detector tensors of the three ET's interferometers to be
\begin{equation}
	\label{eq:det_tensor}
	\begin{split}
		&\boldsymbol{D}_{1} = \frac{1}{2}\left( \boldsymbol{u}\otimes\boldsymbol{u} - \boldsymbol{v}\otimes\boldsymbol{v} \right) \\
		&\boldsymbol{D}_{2} = \frac{1}{2}\left( \boldsymbol{v}\otimes\boldsymbol{v} - \boldsymbol{w}\otimes\boldsymbol{w} \right) \\
		&\boldsymbol{D}_{3} = \frac{1}{2}\left( \boldsymbol{w}\otimes\boldsymbol{w} - \boldsymbol{u}\otimes\boldsymbol{u} \right)
	\end{split}
\end{equation}
where $\boldsymbol{u}$, $\boldsymbol{v}$ and $\boldsymbol{w}$ are the unit vectors along the three arms of ET. One could observe that summing the strain signals from the three detectors cancel the signal regardless of the waveform $h_{m}(t)$ since
\begin{equation}
	\label{eq:signal_sum}
	\begin{split}
		\sum_{n=1}^{3}s_{n}(t) &= \sum_{n=1}^{3}\sum_{m}\Tr\left(\boldsymbol{D}_{n}\boldsymbol{e}_{m}\right)h_{m}(t) \\
		&=\sum_{m}\Tr\left( \left( \sum_{n=1}^{3}\boldsymbol{D}_{n} \right) \boldsymbol{e}_{m}\right)h_{m}(t)
	\end{split}
\end{equation}
but $\sum\limits_{n=1}^{3}\boldsymbol{D}_{n} = \boldsymbol{0}$ as is evident from Eq.~\eqref{eq:det_tensor}, and therefore the sum of strain signals of the three ET's interferometers is zero i.e.\
\begin{equation}
	\label{eq:null_signal}
	\sum_{n=1}^{3}s_{n}(t) = 0
\end{equation}
regardless of the waveform $h_{m}(t)$. The exact cancellation of signal by a linear combination of strain data from a network of detectors is known as \textit{null stream}. In general, the construction of null stream depends on the source location. The construction is however independent of the source location if the detectors are co-located which is the case of ET. Eq.~\eqref{eq:null_signal} could also be interpreted as a linear transformation of the strain signals, and the strain signals are projected onto the null space where there is no GW. This implies that there exists a fixed orthogonal subspace that is the signal space where \textit{all} possible GWs are contained, and we could extract all of the information relevant to GW with a reduced amount of strain data. The generality of Eq.~\eqref{eq:signal_sum} implies that such signal space contains GW with arbitrary polarization content, and Eq.~\eqref{eq:null_signal} holds for GW signals with arbitrary polarization content.

\section{Signal space}
\label{sec:signal_space}

\noindent Since linear algebra is the natural language to discuss linear transformations, we use a matrix formulation of the problem. The projection onto the null space in Eq.~\eqref{eq:null_signal} could be represented by a projection matrix $\boldsymbol{P}_{\text{null}}$ defined by
\begin{equation}
	\boldsymbol{P}_{\text{null}} = 
	\begin{bmatrix}
		1/3 & 1/3 & 1/3 \\
		1/3 & 1/3 & 1/3 \\
		1/3 & 1/3 & 1/3
	\end{bmatrix}
\end{equation}
which we have
\begin{equation}
	\begin{split}
		\boldsymbol{P}_{\text{null}}\boldsymbol{s}(t) &= 
		\begin{bmatrix}
			1/3 & 1/3 & 1/3 \\
			1/3 & 1/3 & 1/3 \\
			1/3 & 1/3 & 1/3
		\end{bmatrix}
		\begin{bmatrix}
			s_{1}(t) \\
			s_{2}(t) \\
			s_{3}(t)
		\end{bmatrix} \\
		&=
		\frac{1}{3}
		\begin{bmatrix}
			s_{1}(t) + s_{2}(t) + s_{3}(t) \\
			s_{1}(t) + s_{2}(t) + s_{3}(t) \\
			s_{1}(t) + s_{2}(t) + s_{3}(t)
		\end{bmatrix} \\
		&=\boldsymbol{0}
	\end{split}
\end{equation}
where $\boldsymbol{s}(t)$ is the strain signal. The factor $\frac{1}{3}$ is added such that $\boldsymbol{P}_{\text{null}}^{2} = \boldsymbol{P}_{\text{null}}$ is satisfied and therefore $\boldsymbol{P}_{\text{null}}$ is a projection matrix. The orthogonal projection which we term as signal projection could be represented by another projection matrix defined by
\begin{equation}
	\label{eq:Pgw}
	\begin{split}
		\boldsymbol{P}_{\text{sig}} &:= \boldsymbol{I} - \boldsymbol{P}_{\text{null}} \\
		&=
		\begin{bmatrix}
			2/3 & -1/3 & -1/3 \\
			-1/3 & 2/3 & -1/3 \\
			-1/3 & -1/3 & 2/3
		\end{bmatrix}
	\end{split}
\end{equation}
where $\boldsymbol{I}$ is a $3\times 3$ identity matrix. $\boldsymbol{P}_{\text{sig}}$ projects the strain data onto the signal space and the data in the null space is removed. Since there is only one linearly independent row vector in $\boldsymbol{P}_{\text{null}}$, and therefore $\boldsymbol{P}_{\text{null}}$ is a rank one matrix. The orthogonal projection matrix $\boldsymbol{P}_{\text{sig}}$ removes one dimension from the strain data $\boldsymbol{d}(t)$, and this implies that there are only two dimensions in the 3-detector strain data that are relevant to GW data analysis. The original 3-detector strain data with $3N$ data points where $N$ is the number of data points in each time series could be compressed to a more compact representation with $2N$ data points without any loss of GW information.

There exist coordinate systems such that the signal space are represented by two coordinate axes and the null space is represented by one coordinate axis. We term such group of coordinate systems as the \textit{principal coordinate system}. Such coordinate system is not unique, but we may follow the following procedures to obtain the corresponding rotation matrix. We first perform the singular value decomposition (SVD) of $\boldsymbol{P}_{\text{sig}}$ and we have
\begin{equation}
	\boldsymbol{P}_{\text{sig}} = \boldsymbol{U}\boldsymbol{S}\boldsymbol{V}^{T}
\end{equation}
where $\boldsymbol{U}$ is a $3\times 3$ unitary matrix, $\boldsymbol{S}$ is a $3\times 3$ diagonal matrix with the singular values on the diagonal, and $\boldsymbol{V}$ is a $3\times 3$ unitary matrix. Since $\boldsymbol{P}_{\text{sig}}$ is symmetric, we have $\boldsymbol{U} = \boldsymbol{V}$. It can also be shown that $\boldsymbol{S} = \text{diag}(1, 1, 0)$. Since there are two degenerate singular values, $\boldsymbol{U}$ and $\boldsymbol{V}$ are not unique. $\boldsymbol{U}\boldsymbol{Q}$ and $\boldsymbol{V}\boldsymbol{Q}$ where $\boldsymbol{Q}$ is an orthogonal matrix that satisfies $\boldsymbol{Q}^{T}\boldsymbol{S}\boldsymbol{Q} = \boldsymbol{S}$ are also solutions. The matrix $\boldsymbol{Q}$ represents the rotation about the null axis. The rotation matrix that aligns the data to the first two axes onto the signal space and the last axis onto the null space is therefore $\boldsymbol{U}^{T}$. $\boldsymbol{U}$ is degenerate with the rotation matrix $\boldsymbol{Q}$ as discussed above, and we could simply choose one of the possible solutions. One possible solution is
\begin{equation}
	\boldsymbol{U} = 
	\begin{bmatrix}
		-\sqrt{6}/6 & -\sqrt{2}/2 & \sqrt{3}/3 \\
		\sqrt{6}/3 & 0 & \sqrt{3}/3 \\
		-\sqrt{6}/6 & \sqrt{2}/2 & \sqrt{3}/3
	\end{bmatrix}\,.
\end{equation}
The details of derivation are shown in Appendix \ref{app:svd_Pgw}.

\section{Data analysis}
\label{sec:pe}

\noindent In this section, we will establish the statistics for performing data analysis in the signal space. We will show that the posterior distribution of the model parameters obtained from the parameter estimation (PE) in the signal space is \textit{identical} to that obtained from the conventional PE using the full set of strain data. 

To facilitate the mathematical expressions, we denote a vector of discrete data of detector $i$ as $\boldsymbol{x}_{i}$, a matrix of discrete data of the three ET interferometers as 
\begin{equation}
	\boldsymbol{x} =
	\begin{bmatrix}
		\boldsymbol{x}_{1}^{T} \\
		\boldsymbol{x}_{2}^{T} \\
		\boldsymbol{x}_{3}^{T}
	\end{bmatrix}\,,
\end{equation}
and the value of the $j$-th entry of the vector of data of detector $i$ as $x_{ij}$.

\subsection{Observation model}
\label{sec:obs_model}

\noindent The detector response of a GW in Eq.~\eqref{eq:det_response}
\begin{equation}
	s(t) = \sum_{m}\sum_{i,j}D_{ij}e_{m}^{ij}h_{m}(t)
\end{equation}
is more commonly written as
\begin{equation}
	s(t) = \sum_{m}F_{m}h_{m}(t)
\end{equation}
where $F_{m} = \sum\limits_{i,j}D_{ij}e_{m}^{ij}$ is called the beam pattern function, and the subscript $m$ labels the polarization mode. We may then formulate the additive-noise observation model of ET with the matrix notations as follows
\begin{equation}
	\label{eq:obs_model}
	\boldsymbol{d}(t) = \boldsymbol{F}\boldsymbol{h}(t) + \boldsymbol{n}(t)
\end{equation}
where $\boldsymbol{d}(t)\in\mathbb{R}^{3}$ is the observed strain data, $\boldsymbol{F}\in\mathbb{R}^{3\times M}$ is the beam pattern function matrix which columns are the beam pattern function of each polarization mode, $\boldsymbol{h}(t)\in\mathbb{R}^{M}$ is the polarization modes, and $\boldsymbol{n}(t)\in\mathbb{R}^{3}$ is the detector noise. Here we do not specify $M$ and the polarization content of the signal for generality. The results that we present do not only apply to the regular tensorial signal model, but also apply to signal models with arbitrary polarization content.

In the principal coordinate system, we have
\begin{equation}
	\boldsymbol{d}^{\text{p}}(t) = \boldsymbol{F}^{\text{p}}\boldsymbol{h}(t) + \boldsymbol{n}^{\text{p}}(t)
\end{equation}
where
\begin{equation}
	\boldsymbol{d}^{\text{p}}(t) = \boldsymbol{U}^{T}\boldsymbol{d}(t) \,,
\end{equation}
\begin{equation}
	\boldsymbol{F}^{\text{p}} = \boldsymbol{U}^{T}\boldsymbol{F} \,,
\end{equation}
and
\begin{equation}
	\boldsymbol{n}^{\text{p}}(t) = \boldsymbol{U}^{T}\boldsymbol{n}(t)\,.
\end{equation}
The signal matrix
\begin{equation}
	\boldsymbol{s}^{\text{p}}(t) = \boldsymbol{U}^{T}\boldsymbol{F}\boldsymbol{h}(t)
\end{equation}
has zero entries in the last row since
\begin{equation}
	\begin{split}
		\boldsymbol{s}^{\text{p}}(t) &= \boldsymbol{U}^{T}\boldsymbol{F}\boldsymbol{h}(t) \\
		&=\boldsymbol{U}^{T}(\boldsymbol{P}_{\text{sig}} + \boldsymbol{P}_{\text{null}})\boldsymbol{F}\boldsymbol{h}(t) \\
		&=\boldsymbol{U}^{T}\boldsymbol{P}_{\text{sig}}\boldsymbol{F}\boldsymbol{h}(t) \\
		&=\boldsymbol{U}^{T}\boldsymbol{U}\boldsymbol{S}\boldsymbol{V}^{T}\boldsymbol{F}\boldsymbol{h}(t) \\
		&=\boldsymbol{S}\boldsymbol{V}^{T}\boldsymbol{F}\boldsymbol{h}(t)
	\end{split}
\end{equation}
and $\boldsymbol{S} = \text{diag}(1, 1, 0)$. Since there is no signal content in the last row, the information of signal could only be extracted from the first two rows of $\boldsymbol{d}^{\text{p}}(t)$. We could truncate the last column of $\boldsymbol{U}$ and define the truncated matrix as $\bar{\boldsymbol{U}}$. $\bar{\boldsymbol{U}}^{T}$ then transforms the data to the GW coordinate system and truncates the component of the data in the null space. The compact representation of data in the signal space is therefore obtained by applying the matrix $\bar{\boldsymbol{U}}^{T}$. The observation model in the signal space could then be written as
\begin{equation}
	\bar{\boldsymbol{d}}^{\text{p}}(t) = \bar{\boldsymbol{F}}^{\text{p}}\boldsymbol{h}(t) + \bar{\boldsymbol{n}}^{\text{p}}(t)
\end{equation}
where
\begin{equation}
	\bar{\boldsymbol{d}}^{\text{p}}(t) = \bar{\boldsymbol{U}}^{T}\boldsymbol{d}(t) \,,
\end{equation}
\begin{equation}
	\bar{\boldsymbol{F}}^{\text{p}} = \bar{\boldsymbol{U}}^{T}\boldsymbol{F} \,,
\end{equation}
and
\begin{equation}
	\bar{\boldsymbol{n}}^{\text{p}}(t) = \bar{\boldsymbol{U}}^{T}\boldsymbol{n}(t)\,.
\end{equation}

\subsection{Likelihood function}

\noindent To perform PE, one would need the likelihood function that describes the noise distribution or more precisely the distribution of data in the signal space given the set of model parameters i.e.\ $p(\bar{\boldsymbol{d}}^{\text{p}}|\boldsymbol{\theta})$ where $\boldsymbol{\theta}$ is the set of model parameters. Since the ET interferometers have the same configurations and are co-located, the noise properties should be similar across detectors. The main noise contribution from thermal or quantum noises above $10$ Hz is not expected to have dependency across detectors, but the extent of noise dependency due to the common displacement at the end stations is yet to be investigated \cite{Regimbau_2012}. For simplicity, we present the likelihood function when the noise follows the zero-mean stationary Gaussian distribution with identical statistical moments across the detectors, and the noises are independent across detectors. The likelihood function of the untransformed data $\boldsymbol{d}$ then takes the form
\begin{widetext}
	\begin{equation}
		p(\boldsymbol{d}|\boldsymbol{\theta}) =
		\prod_{i=1}^{3}\frac{1}{(2\pi)^{N/2}|\boldsymbol{\Sigma}|^{1/2}}\exp\left(-\frac{1}{2}(\boldsymbol{d}_{i}-\boldsymbol{s}_{i}(\boldsymbol{\theta}))^{T}\boldsymbol{\Sigma}^{-1}(\boldsymbol{d}_{i} - \boldsymbol{s}_{i}(\boldsymbol{\theta}))\right)
	\end{equation}
\end{widetext}
where $\boldsymbol{d}_{i}\in\mathbb{R}^{N}$ is the observed strain data at detector $i$, $\boldsymbol{s}_{i}(\boldsymbol{\theta})\in\mathbb{R}^{N}$ is the signal at detector $i$ given the model parameters $\boldsymbol{\theta}$, $\boldsymbol{\Sigma}\in\mathbb{R}^{N\times N}$ is the covariance of noise in each detector, and $N$ is the number of data points in the strain data of each detector. With slightly abused notations, the likelihood function of $\boldsymbol{d}^{\text{p}}$ in the principal coordinate system can be shown to be
\begin{widetext}
	\begin{equation}
		\label{eq:like_signal_space}
		p(\boldsymbol{d}^{p}|\boldsymbol{\theta}) = 
		\prod_{i=1}^{3}\frac{1}{(2\pi)^{N/2}|\boldsymbol{\Sigma}|^{1/2}}\exp\left(-\frac{1}{2}(\boldsymbol{d}_{i}^{\text{p}}-\boldsymbol{s}_{i}^{\text{p}}(\boldsymbol{\theta}))^{T}\boldsymbol{\Sigma}^{-1}(\boldsymbol{d}_{i}^{\text{p}} - \boldsymbol{s}_{i}^{\text{p}}(\boldsymbol{\theta}))\right)\,.
	\end{equation}
\end{widetext}
The details of the derivation could be found in Appendix \ref{app:signal_space_like}. The functional form of Eq.~\eqref{eq:like_signal_space} suggests that the noise components across the coordinate axes are independent of each other, and the covariance of noise in each coordinate is exactly $\boldsymbol{\Sigma}$.

\subsubsection{Time-domain likelihood}

\noindent It can be shown that the corresponding time-domain signal-space likelihood $p(\bar{\boldsymbol{d}}^{\text{p}} | \boldsymbol{\theta})$ takes the form
\begin{widetext}
	\begin{equation}
		\label{eq:gw_space_like_time}
		p(\bar{\boldsymbol{d}}^{\text{p}} | \boldsymbol{\theta}) = 	\prod_{i=1}^{2}\frac{1}{(2\pi)^{N/2}|\boldsymbol{\Sigma}|^{1/2}}\exp\left(-\frac{1}{2}(\bar{\boldsymbol{d}}_{i}^{\text{p}}-\bar{\boldsymbol{s}}_{i}^{\text{p}}(\boldsymbol{\theta}))^{T}\boldsymbol{\Sigma}^{-1}(\bar{\boldsymbol{d}_{i}}^{\text{p}} - \bar{\boldsymbol{s}}_{i}^{\text{p}}(\boldsymbol{\theta}))\right)\,.
	\end{equation}
\end{widetext}
The signal-space likelihood and the standard likelihood are related by a $\boldsymbol{\theta}$-independent factor as follows
\begin{equation}
	\label{eq:like_relation}
	Cp(\bar{\boldsymbol{d}}^{\text{p}} | \boldsymbol{\theta}) = p(\boldsymbol{d}|\boldsymbol{\theta})
\end{equation}
where
\begin{equation}
	C = \frac{1}{(2\pi)^{N/2}|\boldsymbol{\Sigma}|^{1/2}}\exp\left(-\frac{1}{2}
	(\boldsymbol{d}_{3}^{\text{p}})^{T}\boldsymbol{\Sigma}^{-1}\boldsymbol{d}_{3}^{\text{p}}\right)\,.
\end{equation}
The details of the derivation can be found in Appendix \ref{app:signal_space_like}.

\subsubsection{Frequency-domain likelihood}

As the signal-space likelihood takes exactly the same functional form with the standard likelihood in the time domain, the frequency-domain signal-space likelihood immediately follows from the Whittle likelihood and we have
\begin{widetext}
	\begin{equation}
		\label{eq:gw_space_like_freq}
		p(\bar{\boldsymbol{d}}^{\text{p}} | \boldsymbol{\theta}) = \prod_{i=1}^{2}
		\prod_{k=1}^{K}
		\left(
		\frac{2\Delta f}{\pi S[k]}
		\right)
		\exp\left(
		-2\Delta f\frac{\norm{\tilde{\bar{d}}_{ik}^{\text{p}} - \tilde{\bar{s}}_{ik}^{\text{p}}}^{2}}{S[k]}
		\right)\,.
	\end{equation}
\end{widetext}	
where $\tilde{\bar{d}}_{ik}^{\text{p}}$ and $\tilde{\bar{s}}_{ik}^{\text{p}}$ are the $k^{\text{th}}$ bin of the discrete Fourier transform of $\bar{\boldsymbol{d}}_{i}^{\text{p}}$ and $\bar{\boldsymbol{s}}_{i}^{\text{p}}$ respectively, $S[k]$ is the $k^{\text{th}}$ bin of the one-sided power spectral density (PSD) of the noise in each detector, and $\Delta f$ is the frequency resolution of the discrete Fourier transform.

\subsection{Equivalence between the standard PE and the signal-space PE}

\noindent Although the noise in the null space is discarded in the signal-space PE, the noise removal does not improve the PE constraint compared to that of the standard PE using the full set of data. This could be understood by considering the fact that the signal space defines a collection of all possible strain signals projected on a hyperplane spanned by the beam pattern function vectors i.e.\ the column vectors of $\boldsymbol{F}$ in Eq.~\eqref{eq:obs_model}. For the co-located ET detectors, the column vectors always span the same hyperplane for any sky location of the source, and such hyperplane is orthogonal to the null space. The GW waveform models define a collection of polarization modes and the measured strain signals are also the projection of the polarization modes onto the hyperplane. When performing PE using the full set of data, we implicitly fit the data over the hyperplane with the null stream noise being untouched. The removal of null stream noise therefore does not ease the fitting. The posterior distribution of the model parameters using the full set of data is obtained using the Bayes' theorem
\begin{equation}
	p(\boldsymbol{\theta} | \boldsymbol{d}) = \frac{p(\boldsymbol{d} | \boldsymbol{\theta}) p(\boldsymbol{\theta})}{\int p(\boldsymbol{d} | \boldsymbol{\theta}) p(\boldsymbol{\theta}) d\boldsymbol{\theta}}
\end{equation}
where $\boldsymbol{p}(\boldsymbol{\theta})$ is the prior distribution of $\boldsymbol{\theta}$, but since $p(\boldsymbol{d} | \boldsymbol{\theta}) = Cp(\bar{\boldsymbol{d}}^{\text{p}} | \boldsymbol{\theta})$ as stated in Eq.~\eqref{eq:like_relation}, we have
\begin{equation}
	\begin{split}
		p(\boldsymbol{\theta} | \boldsymbol{d}) &= \frac{p(\boldsymbol{d} | \boldsymbol{\theta}) p(\boldsymbol{\theta})}{\int p(\boldsymbol{d} | \boldsymbol{\theta}) p(\boldsymbol{\theta}) d\boldsymbol{\theta}} \\
		&= \frac{Cp(\bar{\boldsymbol{d}}^{\text{p}} | \boldsymbol{\theta}) p(\boldsymbol{\theta})}{\int Cp(\bar{\boldsymbol{d}}^{\text{p}} | \boldsymbol{\theta}) p(\boldsymbol{\theta}) d\boldsymbol{\theta}} \\
		&= \frac{p(\bar{\boldsymbol{d}}^{\text{p}} | \boldsymbol{\theta}) p(\boldsymbol{\theta})}{\int p(\bar{\boldsymbol{d}}^{\text{p}} | \boldsymbol{\theta}) p(\boldsymbol{\theta}) d\boldsymbol{\theta}} \\
		&= p(\boldsymbol{\theta} | \bar{\boldsymbol{d}}^{\text{p}})
	\end{split}
\end{equation}
which proves the equivalence between the posterior distributions of the model parameters using the signal-space data and the full set of data.

\subsection{Invariance of the Bayes factor}

\noindent The Bayes factor between two competing hypotheses $\mathcal{H}_{1}$ and $\mathcal{H}_{2}$ is also invariant after discarding the noise in the null space since
\begin{equation}
	\begin{split}
		\mathcal{B}_{\mathcal{H}_{1}}^{\mathcal{H}_{2}}(\boldsymbol{d}) 
		&=\frac{p(\boldsymbol{d} | \mathcal{H}_{2})}{p(\boldsymbol{d}| \mathcal{H}_{1})}\\
		&=
		\frac{\int p(\boldsymbol{d} | \boldsymbol{\theta};\mathcal{H}_{2})p(\boldsymbol{\theta}; \mathcal{H}_{2})d\boldsymbol{\theta}}
		{\int p(\boldsymbol{d} | \boldsymbol{\theta};\mathcal{H}_{1})p(\boldsymbol{\theta}; \mathcal{H}_{1})d\boldsymbol{\theta}} \\
		&=\frac{\int Cp(\bar{\boldsymbol{d}}^{\text{p}} | \boldsymbol{\theta})p(\boldsymbol{\theta}; \mathcal{H}_{2})d\boldsymbol{\theta}}
		{\int Cp(\bar{\boldsymbol{d}}^{\text{p}} | \boldsymbol{\theta})p(\boldsymbol{\theta}; \mathcal{H}_{1})d\boldsymbol{\theta}} \\
		&=\frac{\int p(\bar{\boldsymbol{d}}^{\text{p}} | \boldsymbol{\theta})p(\boldsymbol{\theta}; \mathcal{H}_{2})d\boldsymbol{\theta}}
		{\int p(\bar{\boldsymbol{d}}^{\text{p}} | \boldsymbol{\theta})p(\boldsymbol{\theta}; \mathcal{H}_{1})d\boldsymbol{\theta}} \\
		&=\frac{p(\bar{\boldsymbol{d}}^{\text{p}} | \mathcal{H}_{2})}{p(\bar{\boldsymbol{d}}^{\text{p}}| \mathcal{H}_{1})}\\
		&=\mathcal{B}_{\mathcal{H}_{1}}^{\mathcal{H}_{2}}(\bar{\boldsymbol{d}}^{\text{p}}) \,.
	\end{split}
\end{equation}
The invariance implies that the distinguishability between competing hypotheses will not be improved after subtracting the null stream noise from the data.

\subsection{Advantages of performing analysis in the signal space}

\noindent Although there is no improvement in the PE constraint and the model distinguishability, the analysis in the signal space involves fewer data since the redundant dimension is discarded. The amount of strain data to be loaded to computer memory is reduced by one-third. If the noise is independent across detectors and shares the same covariance or PSD, the likelihood function of the data in the signal space takes a very simple functional form as shown in Eq.~\eqref{eq:gw_space_like_time} and Eq.~\eqref{eq:gw_space_like_freq}, and the computation of the signal-space likelihood is less expensive than that of the standard likelihood.

\subsection{Unbiased estimation of covariance and PSD of noise using null stream}

\noindent While the GW information can be fully extracted from the data in the signal space, the null space allows us to estimate the noise properties from the signal-free subspace. The covariance or PSD of noise is required in the likelihood function to describe the distribution of data but they are not known exactly, and therefore in practice, we would need to estimate them from the data. The high detection rate of ET \cite{Maggiore_2020} will however pose difficulties for unbiased estimation of the covariance and PSD of noise since it will be more difficult to find a long enough signal-free data segment for the unbiased estimation. This implies that we would need to infer both the signal and the covariance or PSD at the same time when performing PE with the strain data of ET. However, if the assumptions of identical statistical moments and noise independence across detectors are valid, as suggested in Eq.~\eqref{eq:like_signal_space}, the covariance and PSD of the null stream noise are the same as that of the noise in each detector. The existence of the null stream would provide us a signal-free subspace to perform an unbiased estimation of the covariance and PSD of noise in each detector.

\section{Summary and Conclusion}
\label{sec:conclusion}

\noindent We have presented the decomposition of the observation space of ET into the signal space and the null space. The signal space contains all possible GW signals and the null space contains the null stream. We have shown that only the signal-space strain data of the three ET interferometers contain information of the GW source, and therefore one could discard the null stream component from the strain data before performing analysis. We have presented the method to transform the strain data in the original observation space to the signal space.

We have derived the time-domain and frequency-domain likelihood functions of the signal-space data when the detector noise follows the stationary Gaussian distribution with equal covariance among the detectors and the noise is independent across detectors. We have shown that the posterior distribution of the source parameters inferred from the signal-space data is identical to that inferred from the full set of data. We have also shown that the Bayes factor is invariant after discarding the null stream noise which implies that there is no improvement of model distinguishability after discarding the null stream noise from the data. The advantage of performing PE in the signal space is instead the reduced computational cost of the likelihood evaluation, and the reduced computer memory usage to load the data. We also discussed the existence of the null space would allow us to perform an unbiased estimation of covariance and PSD of noise from the signal-free subspace.

The main purpose of the paper is to establish the decomposition of the observation space of ET into the signal space and the null space. We only presented the noise statistics in the signal space when the noise has the same covariance among the detectors and is independent across the detectors. It is however possible for the noise in the three ET detectors to have some dependency since they are colocated. We will leave the investigation of the most general case with unequal covariance and dependency across detectors for future work.

\section{Acknowledgement}

ICFW and TGFL are partially supported by grants from the Research Grants Council of the Hong Kong (Project No. 24\
304317 and 14306419) and Research Committee of the Chinese University of Hong Kong.

\appendix

\section{Singular value decomposition of $\boldsymbol{P}_{\text{sig}}$}
\label{app:svd_Pgw}
\noindent We want to solve the singular value decomposition of $\boldsymbol{P}_{\text{sig}}$
\begin{equation}
	\boldsymbol{P}_{\text{sig}} = \boldsymbol{U}\boldsymbol{S}\boldsymbol{V}^{T}
\end{equation}
where $\boldsymbol{U}$ is a unitary matrix, $\boldsymbol{S}$ is a diagonal matrix with the singular values on the diagonal, , $\boldsymbol{V}$ is a unitary matrix, and
\begin{equation}
	\boldsymbol{P}_{\text{sig}} = 
	\begin{bmatrix}
		2/3 & -1/3 & -1/3 \\
		-1/3 & 2/3 & -1/3 \\
		-1/3 & -1/3 & 2/3
	\end{bmatrix}
\end{equation}
as given in Eq.~\eqref{eq:Pgw}. $\boldsymbol{U}$ and $\boldsymbol{S}$ can be solved by performing the eigenvalue decomposition of $\boldsymbol{P}_{\text{sig}}\boldsymbol{P}_{\text{sig}}^{T}$ since
\begin{equation}
	\label{eq:Pgw_eigen}
	\begin{split}
		&\boldsymbol{P}_{\text{sig}}\boldsymbol{P}_{\text{sig}}^{T} = 
		(\boldsymbol{U}\boldsymbol{S}\boldsymbol{V}^{T})(\boldsymbol{U}\boldsymbol{S}\boldsymbol{V}^{T})^{T} \\
		&\boldsymbol{P}_{\text{sig}} =
		\boldsymbol{U}\boldsymbol{S}^{2}\boldsymbol{U}^{T} \\
		&\boldsymbol{P}_{\text{sig}}\boldsymbol{U} = \boldsymbol{U}\boldsymbol{S}^{2}\,.
	\end{split}
\end{equation}
Therefore, $\boldsymbol{U}$ is a matrix which columns are the eigenvectors of $\boldsymbol{P}_{\text{sig}}$ and $\boldsymbol{S}^{2}$ is a diagonal matrix with the eigenvalues of $\boldsymbol{P}_{\text{sig}}$ on the diagonal. Solving the eigenvalue problem
\begin{equation}
	\boldsymbol{P}_{\text{sig}}\boldsymbol{u} = \lambda\boldsymbol{u} \,,
\end{equation}
and it gives $\lambda = 0 \text{ or }1$. For $\lambda = 0$, the normalized eigenvector is
$\begin{bmatrix}
	\sqrt{3}/3 & \sqrt{3}/3 & \sqrt{3}/3
\end{bmatrix}^{T}
$. For $\lambda = 1$, the orthonormal eigenvectors are
$
\begin{bmatrix}
	-\sqrt{6}/6 & \sqrt{6}/3 & -\sqrt{6}/6
\end{bmatrix}^{T}
$
and
$
\begin{bmatrix}
	-\sqrt{2}/2 & 0 & \sqrt{2}/2
\end{bmatrix}^{T}
$. Hence, we have
\begin{equation}
	\boldsymbol{U} = 
	\begin{bmatrix}
		-\sqrt{6}/6 & -\sqrt{2}/2 & \sqrt{3}/3 \\
		\sqrt{6}/3 & 0 & \sqrt{3}/3 \\
		-\sqrt{6}/6 & \sqrt{2}/2 & \sqrt{3}/3
	\end{bmatrix}
\end{equation}
and
\begin{equation}
	\boldsymbol{S} = 
	\begin{bmatrix}
		1 & 0 & 0 \\
		0 & 1 & 0 \\
		0 & 0 & 0
	\end{bmatrix}\,.
\end{equation}
$\boldsymbol{V}$ can be found by solving the eigenvalue problem of $\boldsymbol{P}_{\text{sig}}^{T}\boldsymbol{P}_{\text{sig}}$ since
\begin{equation}
	\begin{split}
		&\boldsymbol{P}_{\text{sig}}^{T}\boldsymbol{P}_{\text{sig}} =
		(\boldsymbol{U}\boldsymbol{S}\boldsymbol{V}^{T})^{T}(\boldsymbol{U}\boldsymbol{S}\boldsymbol{V}^{T}) \\
		&\boldsymbol{P}_{\text{sig}} = 
		\boldsymbol{V}\boldsymbol{S}^{2}\boldsymbol{V}^{T} \\
		&\boldsymbol{P}_{\text{sig}}\boldsymbol{V} = \boldsymbol{V}\boldsymbol{S}^{2}
	\end{split}
\end{equation}
which is exactly the same eigenvalue problem in Eq.~\eqref{eq:Pgw_eigen}, and we can then conclude $\boldsymbol{V} = \boldsymbol{U}$.

\section{Signal-space likelihood}
\label{app:signal_space_like}

\noindent The strain data $\boldsymbol{d}^{\text{p}}$ and signal $\boldsymbol{s}^{\text{p}}$ in the principal coordinate system are obtained by performing the linear transformation $\boldsymbol{U}^{T}$ as follows
\begin{equation}
	\boldsymbol{d}^{\text{p}} = \boldsymbol{U}^{T}\boldsymbol{d}
\end{equation}
and
\begin{equation}
	\boldsymbol{s}^{\text{p}} = \boldsymbol{U}^{T}\boldsymbol{s}\,.
\end{equation}
The likelihood function $q(\boldsymbol{d}^{\text{p}}|\boldsymbol{\theta})$ is therefore
\begin{equation}
	q(\boldsymbol{d}^{\text{p}}|\boldsymbol{\theta}) = p(\boldsymbol{U}\boldsymbol{d}^{\text{p}}|\boldsymbol{\theta})J_{\boldsymbol{U}}
\end{equation}
where $\boldsymbol{J}_{\boldsymbol{U}}$ is the Jacobian of the linear transformation $\boldsymbol{U}$ and we have $J_{\boldsymbol{U}} = \left|\boldsymbol{U}\right|^{3} = 1$ since $\boldsymbol{U}$ is a unitary matrix. We then have
\begin{widetext}
	\begin{equation}
		\label{eq:d_GW_like}
		\begin{split}
			q(\boldsymbol{d}^{\text{p}}|\boldsymbol{\theta}) &=p(\boldsymbol{U}\boldsymbol{d}^{\text{p}}|\boldsymbol{\theta}) \\
			&=\frac{1}{(2\pi)^{3N/2}|\boldsymbol{\Sigma}|^{3/2}}\exp\left(-\frac{1}{2}\sum_{i=1}^{3}\sum_{j,k=1}^{N}(\boldsymbol{U}\boldsymbol{d}^{\text{p}} - \boldsymbol{U}\boldsymbol{s}^{\text{p}}(\boldsymbol{\theta}))_{ij}(\boldsymbol{\Sigma}^{-1})_{jk}(\boldsymbol{U}\boldsymbol{d}^{\text{p}} - \boldsymbol{U}\boldsymbol{s}^{\text{p}}(\boldsymbol{\theta}))_{ik}
			\right) \\
			&=\frac{1}{(2\pi)^{3N/2}|\boldsymbol{\Sigma}|^{3/2}}
			\exp\left(-\frac{1}{2}\sum_{i,m,n=1}^{3}\sum_{j,k=1}^{N}
			U_{im}(d_{mj}^{\text{p}} - s_{mj}^{\text{p}}(\boldsymbol{\theta}))
			(\boldsymbol{\Sigma}^{-1})_{jk}
			U_{in}(d_{nk}^{\text{p}} - s_{nk}^{\text{p}}(\boldsymbol{\theta}))
			\right) \\
			&=\frac{1}{(2\pi)^{3N/2}|\boldsymbol{\Sigma}|^{3/2}}
			\exp\left(-\frac{1}{2}\sum_{m,n=1}^{3}\sum_{j,k=1}^{N}
			\left(\sum_{i=1}^{3}U_{im}U_{in}\right)
			(d_{mj}^{\text{p}} - s_{mj}^{\text{p}}(\boldsymbol{\theta}))
			(\boldsymbol{\Sigma}^{-1})_{jk}
			(d_{nk}^{\text{p}} - s_{nk}^{\text{p}}(\boldsymbol{\theta}))
			\right) \\
			&=\frac{1}{(2\pi)^{3N/2}|\boldsymbol{\Sigma}|^{3/2}}
			\exp\left(-\frac{1}{2}\sum_{m,n=1}^{3}\sum_{j,k=1}^{N}
			\delta_{mn}
			(d_{mj}^{\text{p}} - s_{mj}^{\text{p}}(\boldsymbol{\theta}))
			(\boldsymbol{\Sigma}^{-1})_{jk}
			(d_{nk}^{\text{p}} - s_{nk}^{\text{p}}(\boldsymbol{\theta}))
			\right) \\
			&=\frac{1}{(2\pi)^{3N/2}|\boldsymbol{\Sigma}|^{3/2}}
			\exp\left(-\frac{1}{2}\sum_{m=1}^{3}\sum_{j,k=1}^{N}
			(d_{mj}^{\text{p}} - s_{mj}^{\text{p}}(\boldsymbol{\theta}))
			(\boldsymbol{\Sigma}^{-1})_{jk}
			(d_{mk}^{\text{p}} - s_{mk}^{\text{p}}(\boldsymbol{\theta}))
			\right) \\
			&=	\prod_{i=1}^{3}\frac{1}{(2\pi)^{N/2}|\boldsymbol{\Sigma}|^{1/2}}\exp\left(-\frac{1}{2}(\boldsymbol{d}_{i}^{\text{p}}-\boldsymbol{s}_{i}^{\text{p}}(\boldsymbol{\theta}))^{T}\boldsymbol{\Sigma}^{-1}(\boldsymbol{d}_{i}^{\text{p}} - \boldsymbol{s}_{i}^{\text{p}}(\boldsymbol{\theta}))\right)
		\end{split}
	\end{equation}
\end{widetext}
where we have used $\sum\limits_{i=1}^{3}U_{im}U_{in} = \delta_{mn}$ in the derivation since $\boldsymbol{U}$ is a unitary matrix. The result suggests that the noises across the coordinates in the principal coordinate system are independent to each other, and the covariance of noise in each coordinate axis is exactly $\boldsymbol{\Sigma}$. The likelihood function in the principal coordinate system can be decomposed into two parts
\begin{widetext}
	\begin{equation}
		\begin{split}
			q(\boldsymbol{d}^{\text{p}}|\boldsymbol{\theta}) &=
			\prod_{i=1}^{3}\frac{1}{(2\pi)^{N/2}|\boldsymbol{\Sigma}|^{1/2}}\exp\left(-\frac{1}{2}(\boldsymbol{d}_{i}^{\text{p}}-\boldsymbol{s}_{i}^{\text{p}}(\boldsymbol{\theta}))^{T}\boldsymbol{\Sigma}^{-1}(\boldsymbol{d}_{i}^{\text{p}} - \boldsymbol{s}_{i}^{\text{p}}(\boldsymbol{\theta}))\right) \\
			&=\left(
			\prod_{i=1}^{2}\frac{1}{(2\pi)^{N/2}|\boldsymbol{\Sigma}|^{1/2}}\exp\left(-\frac{1}{2}(\boldsymbol{d}_{i}^{\text{p}}-\boldsymbol{s}_{i}^{\text{p}}(\boldsymbol{\theta}))^{T}\boldsymbol{\Sigma}^{-1}(\boldsymbol{d}_{i}^{\text{p}} - \boldsymbol{s}_{i}^{\text{p}}(\boldsymbol{\theta}))\right)
			\right)\times \\
			&\qquad\frac{1}{(2\pi)^{N/2}|\boldsymbol{\Sigma}|^{1/2}}\exp\left(-\frac{1}{2}
			(\boldsymbol{d}_{3}^{\text{p}})^{T}\boldsymbol{\Sigma}^{-1}\boldsymbol{d}_{3}^{\text{p}}\right)
		\end{split}
	\end{equation}
\end{widetext}
since $\boldsymbol{s}_{3}^{\text{p}} = \boldsymbol{0}$ as shown in Sec.~\ref{sec:obs_model}. Define $\bar{\boldsymbol{x}}^{\text{p}} = 
\begin{bmatrix}
	(\boldsymbol{x}_{1}^{\text{p}})^{T} \\
	(\boldsymbol{x}_{2}^{\text{p}})^{T}
\end{bmatrix}
$ as the matrix of data in the signal space in the principal coordinate system, we may then define the signal-space likelihood with slightly abused notations to be
\begin{widetext}
	\begin{equation}
		p(\bar{\boldsymbol{d}}^{\text{p}} | \boldsymbol{\theta}) := 
		\prod_{i=1}^{2}\frac{1}{(2\pi)^{N/2}|\boldsymbol{\Sigma}|^{1/2}}\exp\left(-\frac{1}{2}(\boldsymbol{d}_{i}^{\text{p}}-\boldsymbol{s}_{i}^{\text{p}}(\boldsymbol{\theta}))^{T}\boldsymbol{\Sigma}^{-1}(\boldsymbol{d}_{i}^{\text{p}} - \boldsymbol{s}_{i}^{\text{p}}(\boldsymbol{\theta}))\right)
	\end{equation}
\end{widetext}
which describes the distribution of data in the signal space given the model parameters $\boldsymbol{\theta}$. The signal-space likelihood and the standard likelihood therefore differ by a $\boldsymbol{\theta}$-independent factor
\begin{equation}
	Cp(\bar{\boldsymbol{d}}^{\text{p}} | \boldsymbol{\theta}) = p(\boldsymbol{d}|\boldsymbol{\theta})
\end{equation}
where
\begin{equation}
	C = \frac{1}{(2\pi)^{N/2}|\boldsymbol{\Sigma}|^{1/2}}\exp\left(-\frac{1}{2}
	(\boldsymbol{d}_{3}^{\text{p}})^{T}\boldsymbol{\Sigma}^{-1}\boldsymbol{d}_{3}^{\text{p}}\right)\,.
\end{equation}

\clearpage
\bibliography{main}

\end{document}